\documentclass[preprintnumbers,nofootinbib,11pt, showpacs]{revtex4}

\usepackage{bm,amsmath,amssymb}
\usepackage{mathrsfs}
\usepackage[dvips]{graphicx}
\usepackage{color}

\begin{document}

\title{{\Large 
Stable Bound Orbits in Six-dimensional Myers-Perry Black Holes
}}
\author{%
Takahisa Igata
}%
\email{
igata@kwansei.ac.jp
}%

\affiliation{%
Graduate School of Science and Technology, Kwansei Gakuin University, 
Sanda, Hyogo, 669-1337, Japan
}%

\begin{abstract}
The existence of stable bound orbits of test particles is 
one of the most characteristic properties 
in black hole spacetimes. 
In higher-dimensional black holes, due to the dimensionality of gravity, 
there is no stable bound orbit balanced by 
Newtonian gravitational monopole force and centrifugal force 
as in the same mechanism of the four-dimensional Kerr black hole case. 
In this paper, however, the existence of stable bound orbits of 
massive and massless particles 
is shown in six-dimensional singly spinning Myers-Perry black holes with 
a value of the spin parameter larger than a critical value. 
The innermost stable circular orbits and 
the outermost stable bound orbit are found on the rotational axis. 
\end{abstract}


\maketitle

\section{Introduction}
\label{sec:1}
In the last decade, 
higher-dimensional black holes have been studied 
in a broad range of research, 
highly motivated by superstring theories~(see, for example,~\cite{LivingReview}). 
One of the most important black hole solutions in higher-dimensions is 
the Myers-Perry black hole solutions~\cite{Myers:1986un}, 
which are generalization of the four-dimensional Kerr solutions to higher dimensions. 
There also exist the black hole solutions 
that have ring horizon topology in five dimensions,  
called the black ring solutions~\cite{Emparan:2001wn}. 
Many efforts are devoted to understanding physics of these 
black hole solutions.

One of the most important step to comprehend black holes is 
analysis of geodesics. 
Geodesics have been extensively studied 
in various black holes~\cite{Sharp:1979sqa}, 
which provide new insights into not only background geometries
but physical phenomena related to test particles in such backgrounds. 
For example, the geodesic system in the Kerr geometry was studied 
by Carter~\cite{Carter:1968rr}, 
and the Hamilton-Jacobi equation of its system 
was shown to occur the separation of variables by finding the non-trivial constant 
called Carter's constant. 
The discovery of Carter's constant led to 
the understanding of non-trivial hidden symmetries in the Kerr geometry 
such as Killing tensors~\cite{Walker1970}.

In the case that the geodesic equation occurs 
the separation of variables in a black hole spacetime, 
it is possible to study various particle orbits around the black hole analytically. 
Among several special orbits in the Kerr black hole, 
stable bound orbits of massive particles 
are one of the most important characteristics~\cite{Wilkins:1972rs}. 
The stable bound orbits mean that a particle ranges 
over a finite interval of radius, 
neither being captured by the black hole nor escaping to infinity. 
Basically, such orbits as Earth goes around the sun are formed 
by a balance between Newtonian gravitational monopole force 
and centrifugal force in four dimensions. 
Stable bound orbits play important role in the astrophysical context. 
For example, in a black hole and disk system, 
the innermost stable circular orbit 
are thought to be the inner edge of the accretion disk.

In $D$-dimensional black holes in the case $D\geq5$, 
there exist no stable bound orbit balanced 
by Newtonian gravitational monopole force and centrifugal force. 
More precisely, Newtonian gravitational monopole potential term 
and centrifugal potential term in radial motion 
cannot form a potential well because 
the centrifugal potential proportional to $\propto r^{-2}$ in any dimensions 
is dominant in the region far from a black hole 
than the Newtonian gravitational monopole potential 
proportional to $- r^{3-D}$ in general. 
In five-dimensional black rings, however, 
there exist potential wells, which are formed 
in different mechanism~\cite{Igata:2010ye, Igata:2010cd, Igata:2013be}
by contrast to the case in four dimensions,  
and then a particle can be bounded in such a potential well stably. 
This physics is led by having ring horizon topology completely different from round sphere.

In higher-dimensional Myers-Perry black holes, 
horizon shape is deformed by the effect of black hole spin angular momenta. 
Although a horizon takes the shape of almost round sphere 
for small values of spin parameters, in general, 
its shape becomes non-trivial for large values of spin parameters. 
One of the most interesting limit for Myers-Perry black holes is 
the ultra-spinning limit in more than six dimensions~\cite{Emparan:2003sy}. 
Since the horizon is flattened in this limit, 
gravitational field that particles feel changes drastically, and
the existence of stable bound orbits is expected to be 
non-trivial as in the case of five-dimensional black rings.

In $D$-dimensional singly rotating Myers-Perry black holes, 
unstable circular particle orbits was studied in~\cite{Cardoso:2008bp}. 
Althought particle motion has been extensively studied 
in five-dimensional Myers-Perry 
black holes~\cite{Frolov:2003en, Gooding:2008tf, Kagramanova:2012hw, Diemer:2014lba}, 
no stable bound orbit has found yet. 
The purpose of this paper is to show that 
there exist stable bound orbits of particles 
even in higher-dimensional Myers-Perry black holes. 
Note that the mechanism of taking stable bound orbits is physically non-trivial 
as mentioned above. 
In this paper, 
the existence of stable bound orbits is demonstrated 
in six dimensional singly spinning Myers-Perry black holes.

This paper is organized as follows. 
In the following section, 
the geometry of the singly rotating Myers-Perry solution in six dimensions 
is reviewed. 
In Sec.~\ref{sec:3}, 
particle motion orbiting around a six-dimensional singly spinning Myers-Perry black hole 
is analyzed, and the existence of stable bound orbits is 
demonstrated by the analysis of an effective potential. 
In Sec.~\ref{sec:4}, 
the effective potential restricted on the rotational axis 
of a singly spinning Myers-Perry black hole 
is analyzed, 
and the critical value of the spin parameter 
classifying whether or not stable bound orbits exist is determined.
In Sec.~\ref{sec:5}, stable bound orbits of massless particles are 
considered. 
Finally, the implication of our results is discussed in Sec.~\ref{sec:6}.

\section{Singly rotating Myers-Perry black holes in six dimensions}
\label{sec:2}
In this section, 
the geometry of singly rotating Myers-Perry black holes 
in six dimensions is briefly reviewed. 
Its metric in the Boyer-Lindquist coordinates is given by
\begin{align}
ds^2=
&-dt^2
+\frac{\mu}{r\,\Sigma}
\left(dt -a\, \mu_1^2\, d\phi-b \,\mu_2^2\,d\psi\right)^2
+\frac{\Sigma}{\Delta}\,dr^2+r^2 d\alpha^2
\cr
&+\left(r^2+a^2\right)\left(d\mu_1^2+\mu_1^2 \,d\phi^2\right)
+\left(r^2+b^2\right)\left(d\mu_2^2+\mu_2^2 \,d\psi^2\right),
\label{eq:ds^2}
\end{align}
where 
\begin{align}
&\Delta=\frac{\left(r^2+a^2\right)\left(r^2+b^2\right)}{r^2}-\frac{\mu}{r},
\label{eq:Delta}
\\
&\Sigma=\frac{\left(r^2+a^2\right)\left(r^2+b^2\right)}{r^2}
\left(1-\frac{a^2\mu_1^2}{r^2+a^2}-\frac{b^2\mu_2^2}{r^2+b^2}\right),
\label{eq:Sigma}
\\
&\mu_1^2+\mu_2^2+\alpha^2=1,
\label{eq:S^2}
\end{align}
and $\mu$, $a$, and $b$ are the mass and two spin parameters of the family. 
The event horizon is located at the largest real $r$-coordinate value that $\Delta=0$. 
In the case $a=b=0$, 
\eqref{eq:ds^2} reduces to the Schwarzschild-Tangherlini geometry.

\begin{figure}[t]
 \includegraphics[width=6.5cm,clip]{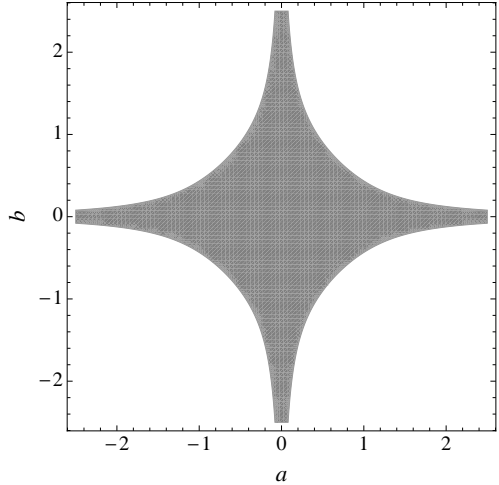}
 \caption{%
 Allowed region of spin parameter space, $a$-$b$ plane, 
 for the regular solutions in the unit $\mu=1$. 
 }%
 \label{fig:a-b}
\end{figure}

Figure~\ref{fig:a-b} shows the allowed region of spin parameters 
that $\Delta=0$ has the real positive roots. 
Note that, in the case of a single rotation, i.e., $a=0$ or $b=0$, 
the allowed region is not bounded along the axes in the figure. 
Indeed, as will be discussed below, 
the spin parameter in the singly rotating case may 
take arbitrarily large value, which is called the ultra-spinning regime.

In order to focus on singly spinning Myers-Perry black holes, 
one of the spin parameters is eliminated, $b=0$, in what follows.
In terms of the following angular coordinates satisfying~\eqref{eq:S^2},
\begin{align}
&\mu_1=\sin \theta,\\
&\mu_2=\cos \theta \sin \chi,\\
&\alpha=\cos \theta \cos \chi,
\end{align}
the metric of the singly rotating Myers-Perry geometry in six dimensions is given by
\begin{align}
ds^2
=-dt^2
+\frac{\mu}{r\,\Sigma}\left(dt-a\sin^2\theta \,d\phi\right)^2
+\frac{\Sigma}{\Delta}\,dr^2
+\Sigma \,d\theta^2
+\left(r^2+a^2\right)\sin^2\theta \,d\phi^2
+r^2\cos^2\theta \,d\Omega^2_2,
\label{eq:ds^2_single}
\end{align}
where 
\begin{align}
d\Omega^2_2=d\chi^2+\sin^2\chi \,d\psi^2. 
\end{align}
Note that \eqref{eq:Delta} and \eqref{eq:Sigma} reduce to 
\begin{align}
&\Delta=r^2+a^2-\frac{\mu}{r},
\\
&\Sigma=r^2+a^2\cos^2\theta. 
\end{align}
Hence, for arbitrary value of $a$, 
there exists at least one real positive root to $\Delta=0$.

As discussed in~\cite{Emparan:2003sy}, 
the boundlessness of the spin parameter allows us to take the limit of large $a$ 
compared to $\mu$, which is called ultra-spinning limit. 
Then \eqref{eq:ds^2_single} near the axis, $\theta\simeq 0$, goes to
\begin{align}
&ds^2=-f\,dt^2+f^{-1}dr^2+r^2\, d\Omega_2^2+d\sigma^2+\sigma^2\,d\phi^2,
\\
&f=1-\frac{\hat{\mu}}{r},
\end{align}
where $\hat{\mu}=\mu/a^2$ kept finite, and 
the new coordinates $\sigma$ is introduced by
\begin{align}
\sigma=a\sin\theta. 
\end{align}
Note that $\sigma$ also is kept finite to obtain the regular metric. 
Hence, the six-dimensional ultra-spinning Myers-Perry black hole geometry 
near the rotational axis 
can be regarded as a direct product space of 
a four-dimensional Schwarzschild black hole 
and a two-dimensional flat space, i.e., a Schwarzschild black brane.

\section{Stable bound orbits for massive particles}
\label{sec:3}
In this section,  massive particle motion around a six-dimensional 
singly spinning Myers-Perry black hole is analyzed. 
Our attention is focused on stable bound orbits, 
which mean that 
a particle ranges over a finite interval of radius outside the horizon, 
neither being captured by the black hole nor escaping to infinity. 
In the method of analyzing an effective potential, 
the existence of stable bound orbits is demonstrated by showing 
explicit shape of effective potentials including a local minimum point. 
Furthermore, the region that stable bound orbits can exist in is also found by 
the analysis of the extremal potential problem.

To reduce the problem of finding stable bound orbits of a particle
to the extremal problem of an effective potential 
in six-dimensional singly rotating Myers-Perry black holes, 
let us define an effective potential in the Hamiltonian formalism.
Let $g^{ab}$ be the inverse metric of \eqref{eq:ds^2_single}. 
The Hamiltonian of particle system in this black holes is given by
\begin{align}
H
&=\frac{1}{2}\left(g^{ab}p_ap_b+\kappa\right)
\\
&=\frac{1}{2}\,\bigg[\,
\frac{\Delta}{\Sigma}\,p_r^2
+\frac{1}{\Sigma}\,p_\theta^2
+\frac{1}{r^2\cos^2\theta}\left(p_\chi^2+\frac{p_\psi^2}{\sin^2\chi}\right)
\cr
&~~~~~~~~~
-\left(1+\frac{\mu\,(r^2+a^2)}{r \,\Delta \Sigma}\right)p_t^2
-\frac{2\,\mu\,a}{r \Delta \Sigma}\,p_t\,p_\phi
+\frac{1}{\Sigma}\left(\frac{1}{\sin^2\theta}
-\frac{a^2}{\Delta}\right)p_\phi^2
+\kappa
\,\bigg],
\label{eq:H}
\end{align}
where $p_a$ are the canonical momenta conjugate to the coordinates, 
and $\kappa$ takes the value of $1$ for massive particles and $0$ for massless particles. 
Note that $p_t$, $p_\phi$, and $p_\psi$ are constants of motion because 
$t$, $\phi$, and $\psi$ are the cyclic coordinates due to the spacetime symmetries. 
Since the relativistic particle system has the reparameterization invariance, 
there exist a constraint, $H=0$, or explicitly in this case, 
\begin{align}
\frac{1}{\Sigma}\left(
\Delta\, p_r^2+p_\theta^2
\right)
+E^2\left(
U+\frac{\kappa}{E^2}
\right)
=0,
\label{eq:constraint}
\end{align}
where $E=-p_t$, and $U$ is defined as 
\begin{align}
U
&=\frac{1}{E^2}
\left(g^{tt}E^2-2g^{t\phi}\,p_\phi E+g^{\phi\phi}\,p_\phi^2+g^{\chi\chi}\,p_\chi^2+g^{\psi\psi}\,p_\psi^2
\right)
\\
&=-1+\frac{\mu\,(r^2+a^2)}{r \Delta \Sigma}
+\frac{2\,\mu\,a}{r \Delta \Sigma}\,\lambda
+\frac{1}{\Sigma}\left(\frac{1}{\sin^2\theta}
-\frac{a^2}{\Delta}\right)\,\lambda^2
+\frac{{\mathscr L}^2}{r^2\cos^2\theta},
\\
{\mathscr L}^2&=\frac{1}{E^2}\left(p_\chi^2+\frac{p_\psi^2}{\sin^2\chi}\right),
\label{eq:L^2}
\end{align}
and $\lambda=p_\phi/E$. 
Note that
${\mathscr L}^2$ is a constant of motion, 
which is Poisson commutable with $p_t$, $p_\phi$, and $p_\psi$, 
and is associated with a reducible Killing tensor.
The function $U$ depends on two variables, $r$ and $\theta$, 
and is called the effective potential in what follows. 
The allowed range of $U$ must be
\begin{align}
-\infty < U\leq -\frac{\kappa}{E^2} \leq 0
\label{eq:U_range}
\end{align}
outside the horizon because the kinetic term, 
the first term in \eqref{eq:constraint}, is non-negative. 
The region that satisfies this inequality shows the allowed region for particle motion.

Let us introduce new coordinates $\zeta$ and $\rho$ instead of $r$ and $\theta$ as
\begin{align}
&r= \sqrt{\frac{\left(\zeta^2+\rho^2-a^2\right)+\Sigma}{2}},
\\
&\sin\theta=\frac{\zeta}{\sqrt{r^2+a^2}},
\\
&\Sigma=\sqrt{(\zeta-a)^2+\rho^2}\sqrt{(\zeta+a)^2+\rho^2},
\end{align}
to grasp effective potential figures intuitively.\footnote{
The new coordinates, $\zeta$ and $\rho$, show  
radial labels of the polar coordinates on each two-dimensional flat planes
because in the flat limit the metric goes to 
\begin{align}
ds^2=-dt^2+d\zeta^2+\zeta^2\,d\psi^2+d\rho^2+\rho^2\,d\phi^2.
\end{align}}
Thus $U$ is transformed into the function of the two variables, 
$\zeta$ and $\rho$, as
\begin{align}
U=U(\zeta, \rho; \,\lambda, {\mathscr L}^2).
\end{align}

The conditions for $U$ to exist stable bound orbits of massive particles 
in six-dimensional singly rotating Myers-Perry black holes are as follows: 
(i) $U$ includes local minimum points; 
(ii) $U$ takes a non-positive value at the local minimum points, 
which comes from \eqref{eq:U_range}. 
Therefore let us analyze the extremal problem of $U$ in what follows. 
The extrema of $U$ are given by
\begin{align}
&\partial_\zeta U=0,
\\
&\partial_\rho U=0.
\end{align}
These equations have a solution $(\zeta, \rho)=(\zeta_0, \rho_0)$, where 
$\zeta_0$ and $\rho_0$ are written by 
$\lambda$ and ${\mathscr L}^2$. 
The expressions for $\zeta_0$ and $\rho_0$ are solved 
in terms of $\lambda$ and ${\mathscr L}^2$ as
\begin{align}
&\lambda=\lambda(\zeta_0, \rho_0)\equiv \lambda_0,
\label{eq:lambda_0}
\\
&{\mathscr L}^2={\mathscr L}^2(\zeta_0, \rho_0) \equiv {\mathscr L}^2_0,
\label{eq:L_0^2}
\end{align}
where $\lambda_0$ is one of the two branches for given $(\zeta_0, \rho_0)$. 
The value of $U$ at an extremum $(\zeta_0, \rho_0)$ is given by
\begin{align}
U_0=U(\zeta_0, \rho_0; \lambda_0, {\mathscr L}^2_{0}).
\end{align}

The next step is to classify whether an extremum of $U$ at $(\zeta_0, \rho_0)$ is 
local minimum or not. 
The second partial derivative test is useful 
to classify an extremal point as local maximum or local minimum, 
and therefore 
let us evaluate the determinant of the Hessian matrix of $U$
\begin{align}
{\mathscr H}(\zeta, \rho; \,\lambda, {\mathscr L}^2)
=\partial_\zeta^2 U\, \partial_\rho^2 U-(\partial_\zeta \partial_\rho U)^2
\end{align}
at each extremal point. 
Let ${\mathscr H}_0$ be the value of ${\mathscr H}$ at $(\zeta_0, \rho_0)$, i.e., 
\begin{align}
{\mathscr H}_0={\mathscr H}(\zeta_0, \rho_0; \lambda_0, {\mathscr L}^2_{0}),
\end{align}
which is the function of $\zeta_0$ and $\rho_0$ 
through \eqref{eq:lambda_0} and \eqref{eq:L_0^2}. 
The second partial derivative test asserts that 
if both $\mathrm{det}{\mathscr H}_0$ and $\mathrm{tr}{\mathscr H}_0$ are positive,
then $(\zeta_0, \rho_0)$ is a local minimum point of $U$.

\begin{figure}[t]
 \includegraphics[width=8.0cm,clip]{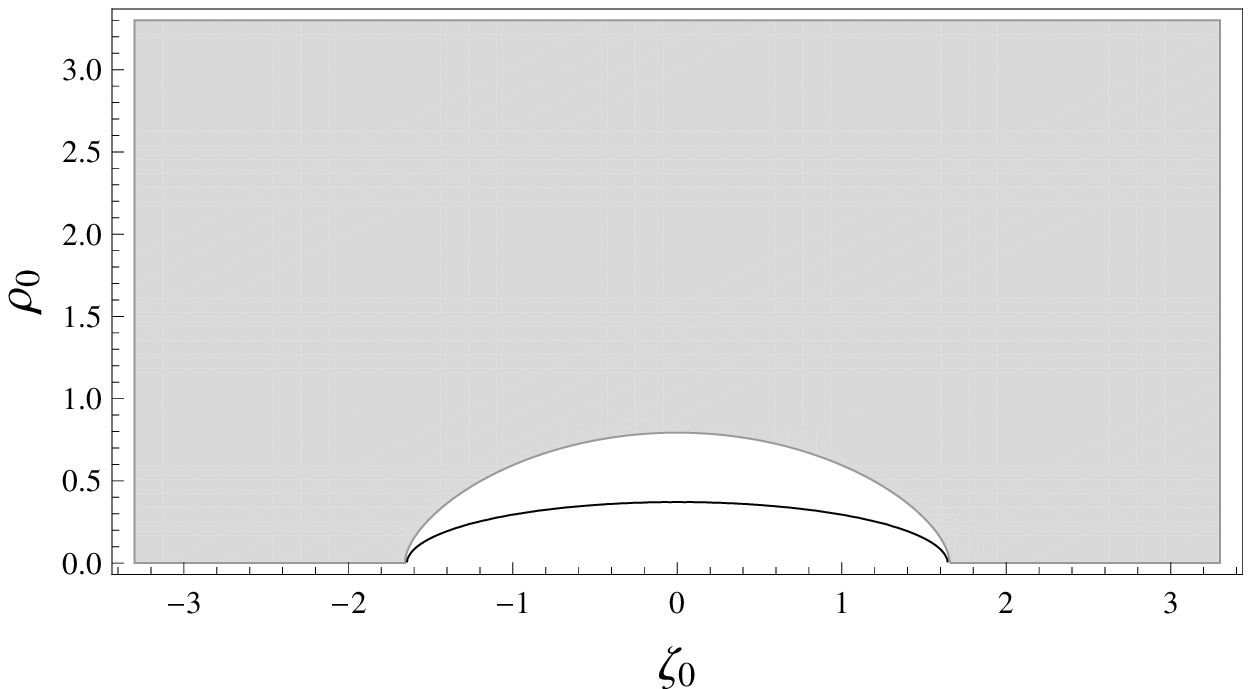}~~~
 \includegraphics[width=8.0cm,clip]{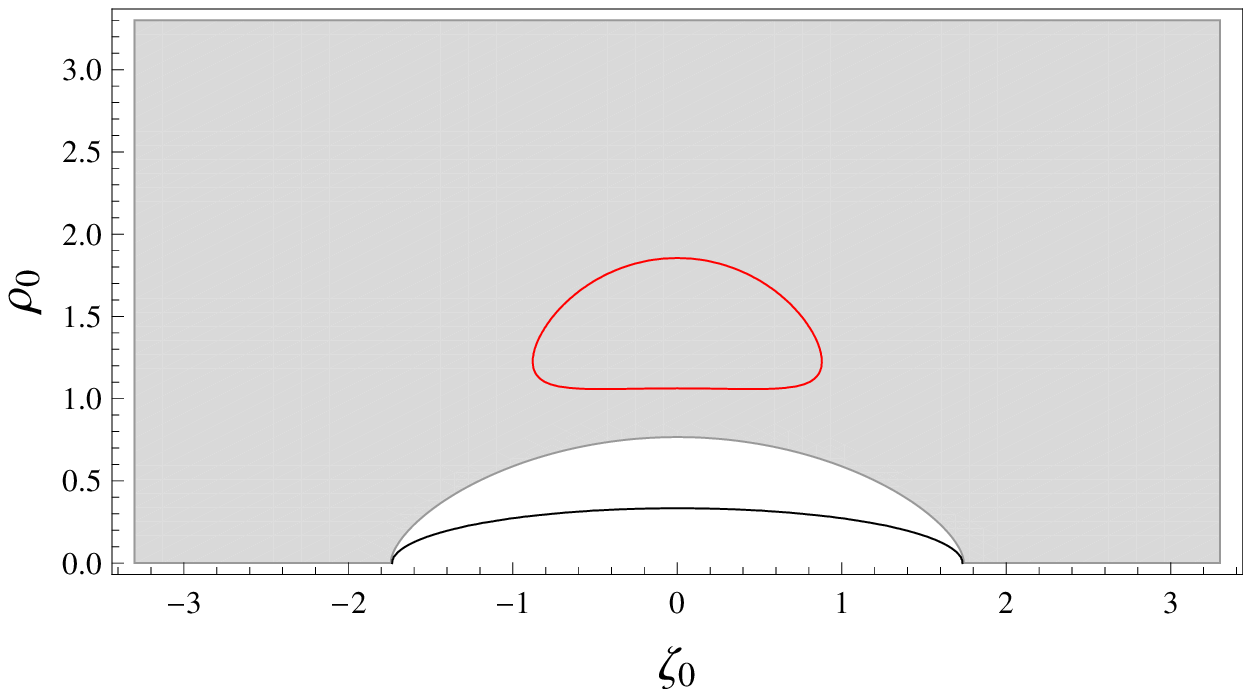}\\[4mm]
 \includegraphics[width=8.0cm,clip]{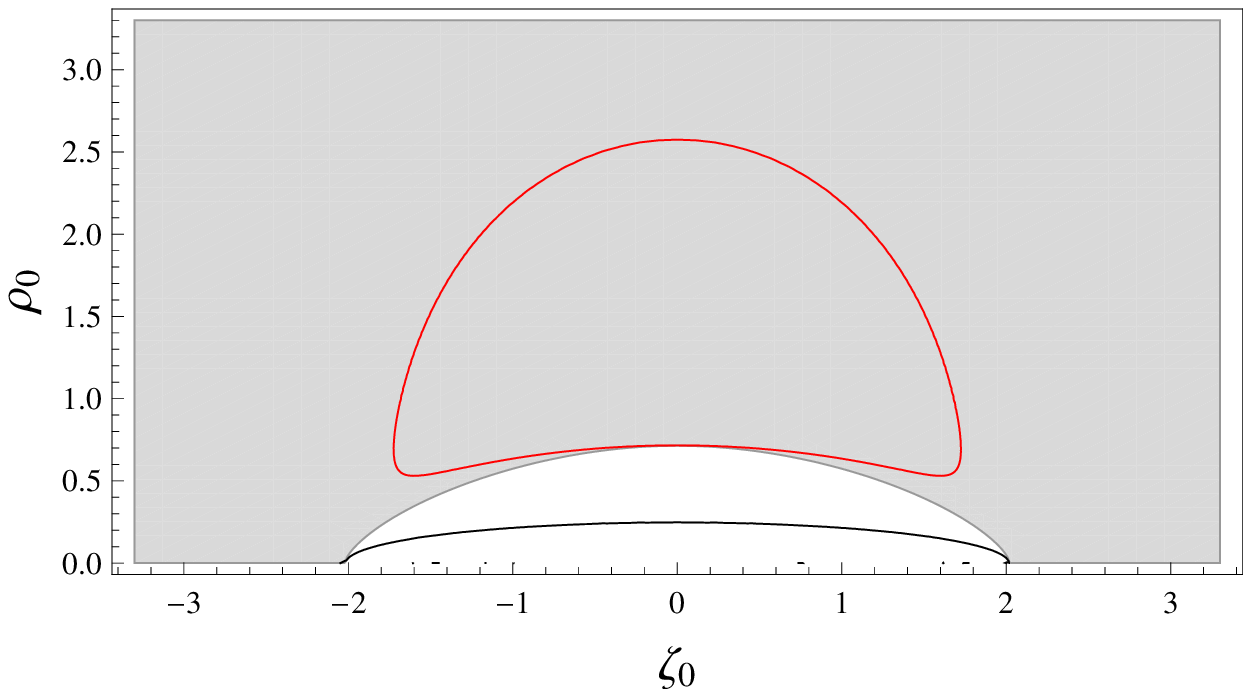}~~~
 \includegraphics[width=8.0cm,clip]{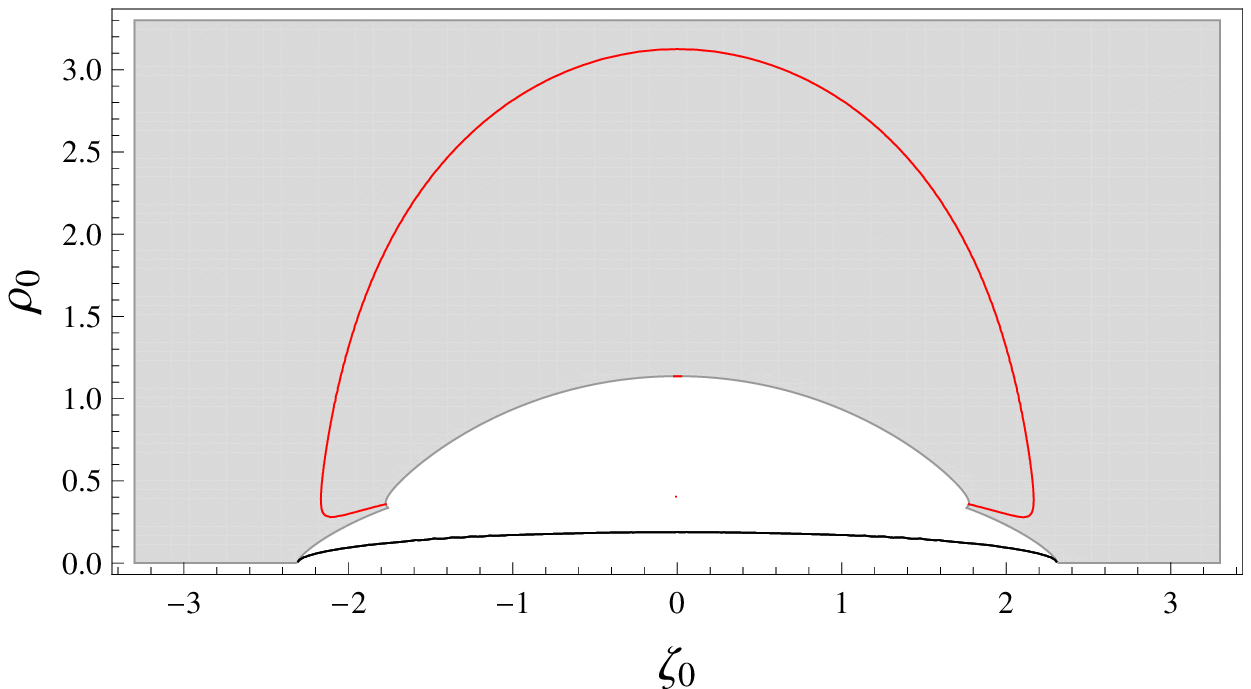}
 \caption{%
 Region where stable bound orbits of massive particles can exist in the
 six-dimensional singly spinning Myers-Perry black holes with $a=1.6$ (upper left panel), 
 $a=1.7$ (upper right panel), $a=2.0$ (lower left panel), and $a=2.3$ (lower right panel), 
 in the unit $\mu=1$. The larger $\lambda_0$-branch is taken here. 
 Each black solid line shows the location of the horizon, 
 and each gray region shows the set satisfying 
 $\lambda_0, {\mathscr L}^2_0 \in \mathbb{R}$, ${\rm tr}{\mathscr H}_0>0$, and $U_0<0$. 
 Inside of each red line shows the set ${\rm det}{\mathscr H}_0>0$. 
 Hence, local minimum points of $U$ can exist in the gray region 
 surrounded by the red lines. 
}%
 \label{fig:massive}
\bigskip
\end{figure}
Then the conditions that stable bound orbits exist in a neighborhood of $(\zeta_0, \rho_0)$ 
are 
\begin{align}
&\lambda_0, {\mathscr L}^2_0 \in \mathbb{R}, 
\\
&\mathrm{det}{\mathscr H}_0>0, 
\\
&\mathrm{tr}{\mathscr H}_0>0,
\\
&U_0 < 0.
\end{align}
Under several fixed values of $a$, 
let us see the region satisfying these conditions. 
The unit is normalized by $\mu$ in the remaining this section. 
The first line in Fig.~\ref{fig:massive} shows that 
local minimum points with non-positive local minimum value 
do appear in the case $a=1.7$ 
although there is no local minimum point of $U$ in the case $a= 1.6$. 
Hence, 
there exist stable bound orbits of massive particles in 
six-dimensional singly spinning Myers-Perry black holes with $a=1.7$,  
and these figures imply that particle motion qualitatively changes 
at a critical value in the range 
$1.6 \leq a \leq 1.7$, 
which will be considered in detail in the following section. 
The second line of Fig.~\ref{fig:massive} shows the case $a=2.0, 2.3$. 
The region that stable bound orbits can exist becomes larger as $a$ gets larger. 
It is worth noting that 
the boundary of the region that stable bound orbits can exist
means marginally stable bound orbits, 
and they include the innermost stable circular orbit and 
the outermost stable circular orbit as spacial cases, 
which will be analyzed in the next section.

\begin{figure}[t]
 \includegraphics[width=8.0cm,clip]{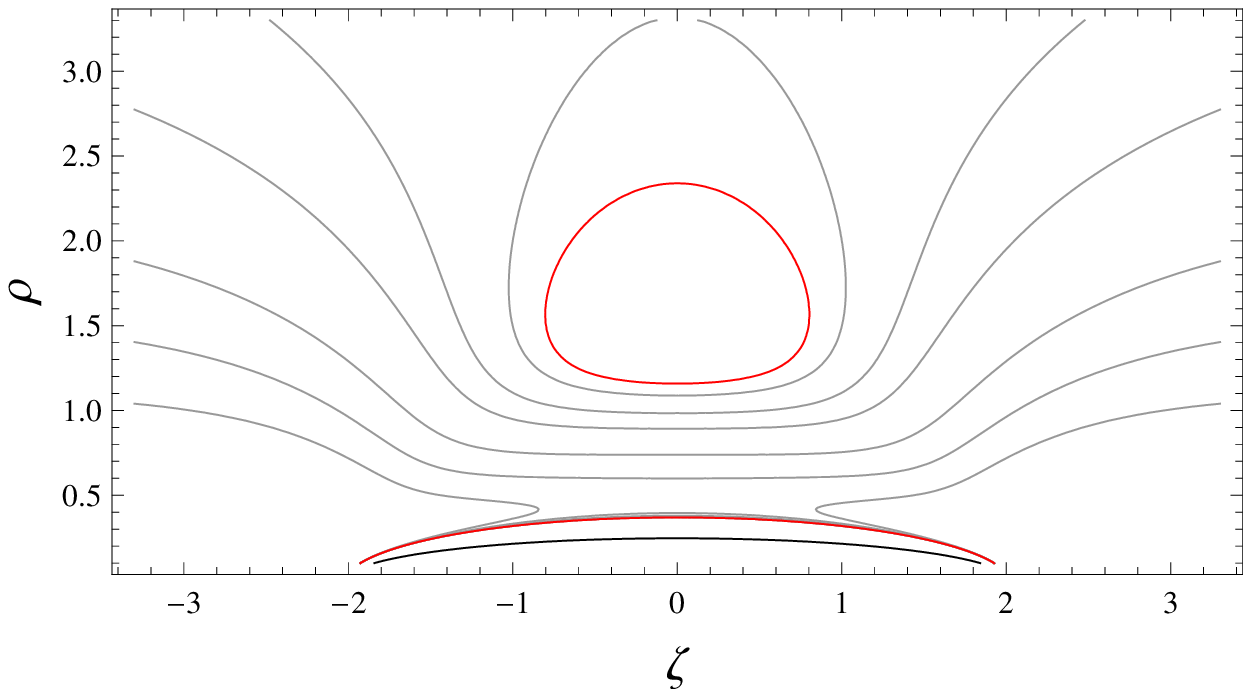}~~
 \includegraphics[width=8.0cm,clip]{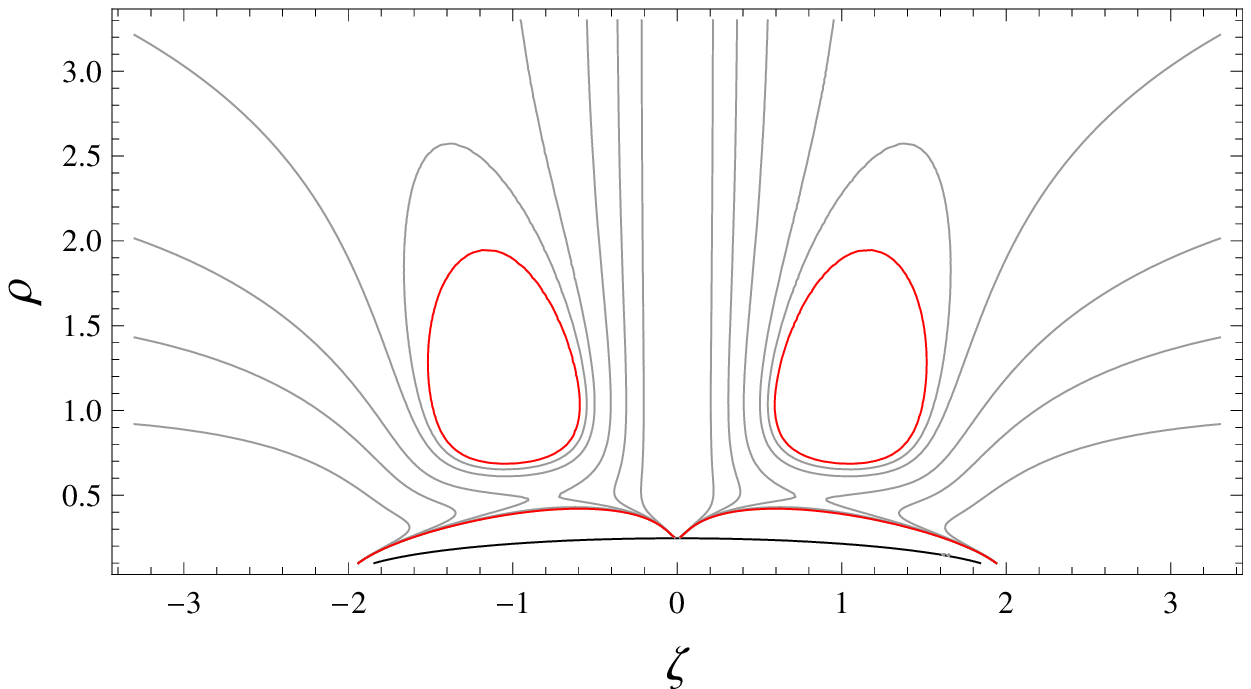}
\caption{%
Explicit shapes of $U$ in the case $a=2$ in the unit $\mu$=1. 
The location of the local minimum point of $U$ is taken at 
$(\zeta_0, \rho_0)=(0, 1.5)$ on the left panel, and 
at $(\zeta_0, \rho_0)=(1, 1)$ on the right panel. 
The solid gray lines show contours of $U$, and 
the black solid lines show the location of the horizon.  
The red solid lines show the contour, $U=-1/E^2$. 
Hence, massive particles can be bounded in the red closed line stably. 
}%
 \label{fig:U_massive}
\bigskip
\end{figure}%
Since Fig.~\ref{fig:massive} provides 
the location of a local minimum point $(\zeta_0, \rho_0)$, 
let us plot $U$ with some local minimum points 
by using the following expression
\begin{align}
U=U(\zeta, \rho; \lambda_0, {\mathscr L}^2_0).
\label{eq:U_massive}
\end{align}
Figure~\ref{fig:U_massive} shows 
explicit shapes of \eqref{eq:U_massive} with a local minimum point 
in the case $a=2$. 
There exist a potential well near the local minimum point.

\section{Critical spin parameter}
\label{sec:4}
In this section, massive particle motion on the rotational axis in 
six-dimensional singly spinning Myers-Perry black holes is analyzed. 
It was indicated in the previous section that 
there was the minimum value of the black hole rotational parameter 
at which the region where the existence of stable bound orbits disappeared, 
which is called the critical parameter in this paper. 
Since this region is located near the rotational axis for rapidly rotating case, 
the value of the critical parameter is expected to be determined by 
analyzing the effective potential on the axis.

First let us define a one-dimensional effective potential 
for particle motion on the rotational axis 
by deriving a set of ordinary differential equations for each variables.  
The Hamilton-Jacobi method is useful for 
the separation of variables of the equation of motion in this case. 
Let $S$ be Hamilton's principal function, and then canonical momenta are 
related to the partial derivatives of $S$ in terms of the conjugate coordinates, 
$p_\mu=\partial S/\partial x^\mu$.  
From \eqref{eq:H} the Hamilton-Jacobi equation is given in the form
\begin{align}
&-\left(1+\frac{\mu\,(r^2+a^2)}{r \,\Delta \Sigma}\right)
\left(\frac{\partial S}{\partial t}\right)^2
-\frac{2a\mu}{r \Delta \Sigma}
\left(\frac{\partial S}{\partial t}\right)
\left(\frac{\partial S}{\partial \phi}\right)
+\frac{1}{\Sigma}\left(\frac{1}{\sin^2\theta}-\frac{a^2}{\Delta}\right)
\left(\frac{\partial S}{\partial \phi}\right)^2
\cr
&+\frac{\Delta}{\Sigma}\left(\frac{\partial S}{\partial r}\right)^2
+\frac{1}{\Sigma}\left(\frac{\partial S}{\partial \theta}\right)^2
+\frac{1}{r^2\cos^2\theta}
\left[
\left(\frac{\partial S}{\partial \chi}\right)^2
+\frac{1}{\sin^2\chi}\left(\frac{\partial S}{\partial \psi}\right)^2
\,\right]
+1=0,
\label{eq:HJeq}
\end{align}
where $\kappa=1$. 
To show the separability of \eqref{eq:HJeq}, 
$S$ is assumed to be 
\begin{align}
S=-E\,t+p_\phi\, \phi+p_\psi \,\psi+S_\chi(\chi)+S_r(r)+S_\theta(\theta),
\label{eq:S}
\end{align}
where $S_\chi (\chi)$, $S_r(r)$, and $S_\theta(\theta)$ 
depend on $\chi$, $r$, and $\theta$, respectively. 
Substitution of \eqref{eq:S} into \eqref{eq:HJeq} leads to 
the separation of variables of the Hamilton-Jacobi equation in the form
\begin{align}
-P(r, \theta)=Q(\chi, \psi)=L^2,
\end{align}
where $L^2$ is the constant separation equivalent to \eqref{eq:L^2}, 
and $P(r, \theta)$, $Q(\chi, \psi)$ are given by
\begin{align}
&P(r,\theta)
=\frac{r^2 \cos^2\theta}{\Sigma}\, \bigg[ 
\left(\frac{d S_\theta}{d \theta}\right)^2
+\frac{p_\phi^2}{\sin^2\theta}
+E^2 a^2\sin^2\theta
+a^2 \cos^2\theta
\cr
&~~~~~~~~~~~~~~~~~~~~~~~~~~
+\Delta \left( \frac{d S_r}{d r} \right)^2
-(r^2+a^2)\left(1+\frac{\mu}{r\Delta}\right)E^2
-\frac{a^2}{\Delta}p_\phi^2
+\frac{2\,\mu\, a}{r\Delta}p_\phi E+r^2\,\bigg],
\\
&Q(\chi, \psi)=\left( \frac{d S_\chi}{d \chi} \right)^2+\frac{p_\psi^2}{\sin^2\chi}, 
\end{align}
respectively. 
Although one of these equation $-P(r,\theta)=L^2$ still depends on the
two variables, $r$ and $\theta$, it
occurs the separation of variable further
\begin{align}
G(\theta)=-F(r)=C^2,
\end{align}
where $C^2$ is the constant separation, 
and $F(r)$, $G(\theta)$ are given by
\begin{align}
&F(r)
=\Delta \left( \frac{d S_r}{d r} \right)^2
-(r^2+a^2)\left(1+\frac{\mu}{r\Delta}\right)E^2
+\frac{2\,\mu\, a}{r\Delta} p_\phi E
-\frac{a^2}{\Delta}p_\phi^2+\frac{a^2}{r^2}L^2
+r^2,
\\
&G(\theta)
=\left(\frac{d S_\theta}{d \theta}\right)^2
+\frac{p_\phi^2}{\sin^2\theta}
+\frac{L^2}{\cos^2\theta}
+E^2 a^2 \sin^2\theta
+a^2 \cos^2\theta,
\end{align}
respectively. 
Note that $C^2$ is associated with 
the non-trivial Killing tensor in the six-dimensional singly rotating Myers-Perry geometry. 
The equations of $r$-motion and $\theta$-motion is written by  
\begin{align}
&\dot r^2+\tilde{V}=0, 
\label{eq:r-motion}
\\
&
\dot \theta^2+\tilde{W}=0,
\end{align}
respectively, where
\begin{align}
&\tilde{V}
=\frac{1}{\Sigma^2}\left(-\left(r^2+a^2\right)^2E^2
+\frac{2\mu a}{r}p_\phi E-a^2 p_\phi^2
+\Delta \left(\frac{a^2}{r^2}L^2+r^2\right)
+\Delta\, C^2\right),
\label{eq:Vtilde}
\\
&\tilde{W}
=\frac{1}{\Sigma^2}\left(
\frac{p_\phi^2}{\sin^2\theta}
+\frac{L^2}{\cos^2\theta}
-C^2
+a^2 \cos^2\theta
+E^2 a^2 \sin^2\theta
\right).
\end{align}
In what follows, $\tilde{V}$ is called the one-dimensional effective potential in $r$-motion, 
and $\tilde{W}$ is called the one-dimensional effective potential in $\theta$-motion.

The next step is the derivation of the one-dimensional effective potential 
in $r$-motion restricted on the rotational axis, $\theta=0$. 
The particle with the initial conditions, $\theta(0)=0$ and $\dot{\theta}(0)=0$, 
must satisfy
\begin{align}
&p_\phi=0,
\label{eq:p_phi=0}
\\
&C^2=L^2+a^2.
\label{eq:C^2=L^2+a^2}
\end{align}
These relations lead to $\ddot{\theta}(0)=0$, 
which means that the particle must keep 
moving on the rotational axis. 
Substitution of \eqref{eq:p_phi=0} 
and \eqref{eq:C^2=L^2+a^2} into \eqref{eq:Vtilde} yields
\begin{align}
\tilde{V}=-\left(E^2-1\right)+\frac{L^2}{r^2}-\frac{\mu\left(r^2+L^2\right)}{r^3\left(r^2+a^2\right)}
\equiv V(r; E, L).
\end{align}
This is the one-dimensional effective potential 
for the particle moving on the axis $\theta=0$.

Let us determine the critical parameter 
by analyzing the extremal value problem of $V$. 
If $r=r_{\rm m}$ is the radii of marginally stable circular orbits, 
the following conditions must be satisfied
\begin{align}
&V(r_{\rm m}; E, L)=0,
\label{eq:Vm}
\\
&V'(r_{\rm m}; E, L)=0.
\label{eq:V'm}
\end{align}
These equations are solved in terms of $E$ and $L$ 
as $E=E_{\rm m}$ and $L=L_{\rm m}$, where
\begin{align}
&E_{\rm m}=\sqrt{\frac{2\left(r_{\rm m}^3+a^2\,r_{\rm m}-\mu\right)^2}{r_{\rm m}
\left(
2\,r_{\rm m}\left(r_{\rm m}^2+a^2\right)^2-\mu\,(5\, r_{\rm m}^2+3\,a^2)
\right)}}, 
\\
&L_{\rm m}^2=\frac{\mu\,r_{\rm m}^2\,(3r_{\rm m}^2+a^2)}{2r_{\rm m}\,(r_{\rm m}^2+a^2)^2-\mu\,(5r_{\rm m}^2+3a^2)}.
\end{align}
In addition to \eqref{eq:Vm} and \eqref{eq:V'm} 
the marginal condition requires
\begin{align}
V''(r_{\rm m}; E_{\rm m}, L_{\rm m})
=\frac{2\,\mu}{r_{\rm m}^3}\frac{
a^4\,r_{\rm m}+3\,a^2\left(2\,r_{\rm m}^3-\mu\right)-3\left(r_{\rm m}^5+5\,\mu\, r_{\rm m}^2\right)
}{2\,r_{\rm m}\left(r_{\rm m}^2+a^2\right)^3-\mu\,\left(r_{\rm m}^2+a^2\right)\left(5\,r_{\rm m}^2+3\,a^2\right)
}=0.
\end{align}
The solution to this equation for $a$ is $a=a_{\rm m}$, where $a_{\rm m}$ is of the form
\begin{align}
a_{\rm m}=\sqrt{\frac{3\,\mu-6\,r_{\rm m}^3+\sqrt{3\left(3\,\mu^2+8\,\mu\, r_{\rm m}^3+16\,r_{\rm m}^6\right)}}{2\,r_{\rm m}}}.
\end{align}
\begin{figure}[t]
 \includegraphics[width=7cm,clip]{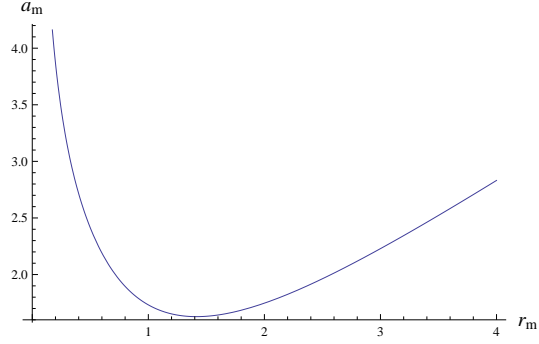}
 \caption{%
Radii of the marginally stable circular orbits for each value of $a$ in the unit $\mu=1$. 
}%
 \label{fig:MSCO}
\bigskip
\end{figure}%
Figure~\ref{fig:MSCO} shows the radii of 
the marginally stable circular orbits for each value of $a$, which are  
the innermost stable circular orbits and the outermost stable circular orbits. 
The result shows that 
$a_{\rm m}$ has the minimum value $a_*$, 
which corresponds to the critical value of the spinning parameter 
as mentioned in Sec.~\ref{sec:3}. 
Finally, the value of $a_*$
is exactly determined as
\begin{align}
a_*=\left(
\frac{15 \sqrt{2\sqrt{10}-5}}{4}\,\mu
\right)^{1/3},
\end{align}
for which $r_{\rm m}$ takes the value $r_*$ such that 
\begin{align}
r_*=\left(\frac{5+\sqrt{10}}{4}\,\mu \right)^{1/3}. 
\end{align}
The value of $a_*$ is approximated as 
\begin{align}
a_*/\mu^{1/3}=1.628\cdots.  
\end{align}
This result implies that stable bound orbits exist for $a\geq a_*$ and disappear at $a< a_*$ 
as expected from the analysis of $U$ in the previous section.

\section{Stable bound orbits for massless particles}
\label{sec:5}
In this section let us consider stable bound orbits of massless particles 
in six-dimensional singly spinning Myers-Perry black holes. 
Although the analysis is not performed systematically as demonstrated for massive particles 
in Sec.~\ref{sec:3} and \ref{sec:4}, 
the evidence for the existence of such orbits is presented 
by showing the suitable shape of $U$. 
\begin{figure}[t]
 \includegraphics[width=8.0cm,clip]{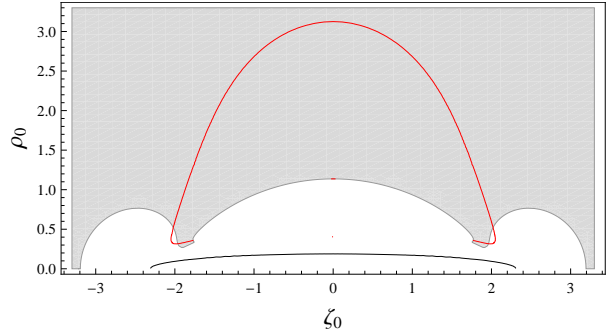}
 \caption{%
 Region where stable bound orbits of massive particles can exist 
 in the case $a=2.3$ in the unit $\mu=1$. 
 Another branch of $\lambda_0$ is chosen, as compared with Fig.~\ref{fig:massive}. 
 The black solid line shows the location of the horizon, 
the gray region shows the set satisfying 
$\lambda_0, {\mathscr L}^2_0 \in \mathbb{R}$, ${\rm tr}{\mathscr H}_0>0$, $U_0<0$. 
Local minimum points of $U$ exist in the gray region surrounded by the red solid line. 
}
 \label{fig:massive-}
\bigskip
\end{figure}%

\begin{figure}[t]
 \includegraphics[width=6.5cm,clip]{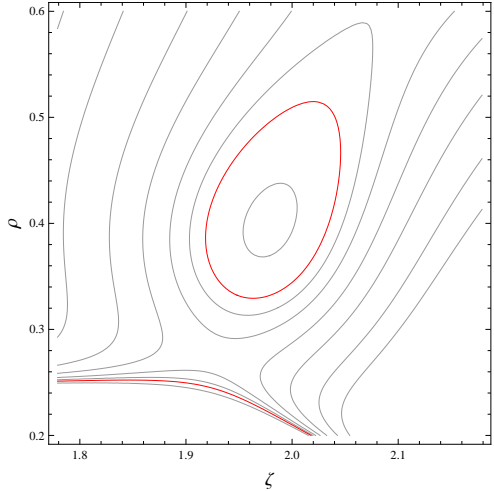}
 \caption{%
Shape of the effective potential $U$ in the case $a=2.3$ in the $\mu=1$ unit. 
The local minimum point of the potential is chosen at 
$(\zeta_0, \rho_0)=(1.979, 0.4)$. 
The gray lines show contours of the potential, and
the red line shows the contour of $U=0$. 
}%
\label{fig:massless}
\bigskip
\end{figure}

Figure~\ref{fig:massive-} shows the region 
where stable bound orbits of massive particles can exist 
in the case $a/\mu^{1/3}=2.3$. 
Note that the other branch for $\lambda_0$ is chosen, as compared with Sec.~\ref{sec:3}. 
It is found that 
the set of marginally stable bound orbits, the red line, 
intersects the boundary of the region satisfying 
$\lambda_0, {\mathscr L}^2_0 \in \mathbb{R}$, ${\rm tr}{\mathscr H}_0>0$, and $U_0<0$, 
the gray region. 
It is worth noting that the condition $U_0=0$ 
determines this part of the boundary.  
Hence, 
massless particles can be marginally stably bounded at this point.  
Furthermore, 
near the marginally stable bound orbit of massless particles, 
there can be found massless particles stably bounded inside a potential well. 
Indeed, 
Fig.~\ref{fig:massless} shows that an explicit form of $U$ in the case 
that a massless particle is stably bounded in a finite region. 
The allowed region for massless particles is restricted in the red solid closed line, 
and this result shows the existence of stable bound orbits of massless particles.

\section{Conclusion}
\label{sec:6}
In this paper, 
particle motion in higher-dimensional black holes has been investigated to 
understand fundamental properties of higher-dimensional gravity. 
Since higher-dimensional black holes in more than six dimensions 
are allowed to take arbitrarily large spin parameter, 
its characteristic feature of gravitational field 
leads to non-trivial particle dynamics. 
The main result of this paper is to show the existence of stable bound orbits of massive 
and massless particles in six-dimensional singly spinning Myers-Perry black holes. 
In the case that the spin parameter is larger than the critical value, 
massive particles can be stably bounded near rotational axis,
and the region for the existence of stable bound orbits becomes finite. 
As special cases, there exist 
the innermost stable circular orbit and the outermost stable circular orbit 
on the rotational axis. 
On the other hand, 
there is no stable bound orbits for small value of the spin parameter. 
Therefore, behavior of particle motion 
around a six-dimensional singly rotating Myers-Perry black hole qualitatively changes 
at the critical value of the spin parameter $a_*$, which is determined as 
$a_*/\mu^{1/3}=1.628\cdots$.   
It has been also shown that 
massless particles can be also stably bounded 
by six-dimensional 
singly spinning Myers-Perry black holes. 
Since massless particles are not stably bounded 
outside the horizon in four-dimensional Kerr black holes, 
this phenomenon in higher-dimensions is unusual.

As mentioned in Sec.~\ref{sec:2}, 
the boundlessness of the spin parameter 
allows to take 
ultraspinning limit of 
a singly rotating Myers-Perry black hole in six dimensions, 
and then the geometry goes to
a direct product spacetime of a four-dimensional Schwarzschild black hole and 
a two-dimensional flat space at least near rotational axis. 
Therefore, a particle moving near the rotational axis effectively feels 
gravity of the Schwarzschild black brane geometry. 
The appearance of stable bound orbits of massive particles 
is naively understood as in the same mechanism 
of the four-dimensional Schwarzschild black hole case. 
Futhermore, the appearance of stable bound orbits of massless particles 
is also understood in the same way because 
non-zero particle momentum in the residual two-dimensional flat direction 
leads to an effective particle mass.

In the sense of the ultraspinning limit of Myers-Perry black hole in $D(\geq 6)$-dimensions, 
limiting horizon topology is $S^{D-2} \to {\mathbb R}^{2n}\times S^{D-2(n+1)}$, where 
$n$ is the number of larger spin parameters than $N-n$ residual spin parameters and 
$N=\lfloor (D-1)/2 \rfloor$ is the total number of spin parameters~\cite{LivingReview}. 
In the case $n=(D-4)/2$, which is satisfied only in even dimensions, 
the limiting horizon topology includes $S^2$. 
Hence, 
it is conjectured that there exist stable bound orbits of massive 
and massless particles at least in even-dimensional Myers-Perry black holes 
in more than six dimensions.

The existence of stable bound orbits for massless particles 
means that photons and gravitons can be stably bounded 
near a singly spinning Myers-Perry black hole. 
This leads to new physical phenomena or 
implies a sort of instability of the geometry. 
In this sense, the relation between stable bound orbits of massless particles 
and the Gregory-Laflamme instability~\cite{Gregory:1993vy} is interesting issue. 
Not only particle dynamics in six-dimensional singly spinning Myers-Perry black holes 
but field dynamics is also interesting issue for future work.

\subsection*{Acknowledgements}
This work is partially supported by the JSPS Strategic Young Researcher Overseas Visits Program for Accelerating Brain Circulation "Deepening and Evolution of Mathematics and Physics, Building of International Network Hub based on OCAMI", and is partially supported by a Research Grant from the Tokyo Institute of Technology Foundation.

\newpage



\end{document}